\documentclass[11pt]{article}
 
 \usepackage[iso-8859-7]{inputenc}
 \usepackage{epsfig}
 \usepackage{graphicx}
 \usepackage{epstopdf}
\usepackage{amsfonts}
 \usepackage{amssymb}
 \usepackage{amsthm}
 \usepackage{amsmath}
 \usepackage{multirow}
 
 \usepackage{color}

 \newcommand{\beq}{\begin{equation}}
 \newcommand{\eeq}{\end{equation}}
 \newcommand{\bea}{\begin{eqnarray}}
 \newcommand{\eea}{\end{eqnarray}}
 
 \newcommand{\gsim}{\lower.7ex\hbox{$\;\stackrel{\textstyle>}{\sim}\;$}}
 \newcommand{\lsim}{\lower.7ex\hbox{$\;\stackrel{\textstyle<}{\sim}\;$}}

 \newcommand{\be}{\begin{equation}}
 \newcommand{\ee}{\end{equation}}
 \newcommand{\ba}{\begin{eqnarray}}
 \newcommand{\ea}{\end{eqnarray}}

 \newcommand{\ov}{\overline}
 \newcommand{\V}{\mathcal{V}}
 \newcommand{\K}{\mathcal{K}}
 \newcommand{\LN}{\textrm{ln}}
 
 \usepackage{float}
 \usepackage{amsmath} 
 \usepackage{amssymb}   
 \usepackage{amsthm} 
 \usepackage{geometry}
 \usepackage{graphicx} 
 \usepackage{multirow} 
\begin{document} 
 	\thispagestyle{empty}
 	
 	\begin{titlepage}
 		
 		\vspace*{0.7cm}

 		\begin{center}
 			{\Large {\bf Inflation from the internal volume in type IIB/F-theory compactification
			}}
 			\\[12mm]
 			Ignatios Antoniadis$^{a,b}$~\footnote{E-mail: \texttt{antoniad@lpthe.jussieu.fr}}, 
 			Yifan Chen$^{a}$~\footnote{E-mail: \texttt{yifan.chen@lpthe.jussieu.fr}},
 			George K. Leontaris$^{c}$~\footnote{E-mail: \texttt{leonta@uoi.gr}}
 		\end{center}
 		\vspace*{0.50cm}
 		\centerline{$^{a}$ \it
 			Laboratoire de Physique Th\'eorique et Hautes \'Energies - LPTHE,
 		}
 		\centerline{\it
 			Sorbonne Universit\'e, CNRS, 4 Place Jussieu, 75005 Paris, France}
 		\vspace*{0.2cm}
 		\centerline{$^{b}$ \it
 			Albert Einstein Center, Institute for Theoretical Physics, University of Bern,}
 		\centerline{\it
 			Sidlerstrasse 5, CH-3012 Bern, Switzerland}
 		\vspace*{0.2cm}
 		\centerline{$^{c}$ \it
 			Physics Department, University of Ioannina}
 		\centerline{\it 45110, Ioannina, 	Greece}
 		\vspace*{1.20cm}
 \begin{abstract}
We study cosmological inflation within a recently proposed framework of perturbative moduli stabilisation in type IIB/F theory compactifications on Calabi-Yau threefolds. The stabilisation mechanism utilises three stacks of magnetised 7-branes and relies on perturbative corrections to the K\"ahler potential that grow logarithmically in the transverse sizes of co-dimension two due to local tadpoles of closed string states in the bulk. The inflaton is the K\"ahler modulus associated with the internal compactification volume that starts rolling down the scalar potential from an initial condition around its maximum. Although the parameter space allows moduli stabilisation in de Sitter space, the resulting number of e-foldings is too low. An extra uplifting source of the vacuum energy is then required to achieve phenomenologically viable inflation and a positive (although tiny) vacuum energy at the minimum. We discuss a class of uplifting potentials arising from strongly coupled matter fields. In a particular case, they reproduce the effect of the new Fayet-Iliopoulos term  recently discussed in a supergravity context, that can be written for a non R-symmetry $U(1)$ and is gauge invariant at the Lagrangian level.

\end{abstract}
\end{titlepage}

 	\section{Introduction}

 	One of the key challenges in string theory cosmology is to implement a successful scenario	of cosmological inflation with robust predictions of the slow-roll parameters.	There are several distinct ways to realise cosmological inflation in string theory~\cite{Kallosh:2007ig}.  In a wide class of string inflationary models the r\^ole of the inflaton is played by  one of the numerous moduli fields emerging in Calabi-Yau compactifications. These  are  the real or the imaginary part of complex structure and K\"ahler moduli,  or those associated with the relative position 	 of D-branes.
 	In order to realise inflation with these candidates, however, firstly it is crucial  to  ensure that in the effective supergravity theory limit  these moduli fields obtain  the required properties. Of particular importance are the issues of moduli stabilisation and the specific constraints that should be respected by the effective potential.  Thus, at the minimum of the effective potential all  moduli fields should acquire  positive masses squared; also,
 	in accordance with cosmological observations confirming  the accelerated expansion of the universe, a positive but tiny cosmological constant would  be the  optimal solution. 
 	
 	In the context of Calabi-Yau compactifications  with background fluxes, a considerable amount of work has been devoted to constructing  de Sitter (dS) vacua and   stabilising  the moduli fields~\cite{Kachru:2003aw, Balasubramanian:2005zx}. The solution of these major issues~\footnote{There is considerable ongoing debade and conjectures on the  existence of  dS vacua in string theory. For an incomplete list see for example~\cite{Danielsson:2018ztv}. For related 	cosmological implications  and other remarks on the conjectures see also~\cite{Agrawal:2018own}.} are  prerequisites for a successful model of string inflation. In a wide class of models in the context of type II-B string theory, the inflaton field is one of  the K\"ahler moduli  which  appear in the tree-level no-scale type  K\"ahler potential through the compactification volume \cite{Conlon:2005jm,Cicoli:2008gp}. Their masses remain undetermined because they are absent at the tree-level superpotential.  It has been suggested that they can be stabilised through non-perturbative superpotential contributions generated by gaugino condensation in some gauge group factor(s).  In addition, if leading order $\alpha'$ corrections are  included in the K\"ahler potential they break the no-scale form and stabilise the volume. Additional contributions form D-terms or $\overline{D3}$ branes are usually necessary to uplift the vacuum energy and generate a small positive cosmological constant.

 	In a previous paper~\cite{Antoniadis:2018hqy},  working in the framework of type IIB/F-theory we have investigated the possibility of obtaining dS minima and stabilising the K\"ahler moduli, invoking only  perturbative contributions to the K\"ahler potential. More precisely,  we have shown that these two merits are naturally feasible  by relying  on  configurations of at least three  magnetised  spacetime filling 7-branes intersecting each other.   We have argued that  the K\"ahler potential receives  loop corrections  which display in general
 	a logarithmic dependence on the  volume of dimension two,  transverse to each 7-brane~\cite{Antoniadis:1998ax}.
 Moreover, magnetised 7-branes are associated with anomalous abelian symmetries which induce world-volume dependent Fayet-Iliopoulos (FI) terms~\cite{Fayet:1974jb}.  These lead to positive, moduli dependent,  D-term contributions to the scalar potential which  act as an uplifting  mechanism giving rise to a dS minimum in a natural way.

 	In the present work,  we  examine the possibility to implement a successful cosmological inflationary scenario in the above framework. A decisive factor 
 	to achieve this goal is  the specific structure of the effective potential of the moduli fields. 
It turns out that the region of parameter space leading to a dS minimum is too narrow and does not allow to obtain a successful slow-roll cosmological inflation because the number of e-foldings is too low.
Thus, it is of  crucial importance towards a successful prediction of the  inflationary observables to implement the mechanism of perturbative moduli stabilisation with  an extra uplift of the vacuum energy.
 Here, we explore  the consequences of a new class of FI terms in global and local supersymmetry that were proposed recently~\cite{Cribiori:2017laj,Antoniadis:2018cpq, Antoniadis:2018oeh}. Their novelty is that they are gauge invariant at the Lagrangian level and, as such, can be written for any $U(1)$ that does not have to be an R-symmetry as in standard supergravity; they are however singular (non-local) in the supersymmetric limit. Also, they may  involve arbitrary functions of chiral matter superfields; in the simlplest case of no new functions, there are two possibilities: one that gives a contribution to the scalar potential as the one of $\overline{D3}$ branes but violates invariance under K\"ahler transformations~\cite{Cribiori:2017laj, Antoniadis:2018cpq} and another that preserves it and amounts to a constant FI term as the one of global supersymmetry~\cite{Antoniadis:2018oeh}. Here, as an example, we consider the second path.

 	The paper is organised as follows. In Section 2, we review  the salient features of the geometric stabilisation mechanism in the presence of fluxes and intersecting D7-branes. We describe the effective scalar potential which comprises both F- and D-term contributions; in particular, in section 2.2, we  define a canonically normalised basis for the K\"ahler moduli, we identify the volume with the inflaton field, and compute the mass spectrum around the minimum of the scalar potential; in subsection 2.3, we determine the parameter space for a dS minimum and study the prospects of inflation, starting with initial conditions around the maximum. We find that although the slow-roll conditions can be satisfied, the number of e-foldings are too low in the parameter region where the minimum of the potential has positive energy. An extra uplifting term is then required. In Section 3, we discuss the new FI terms that were proposed recently in a supergravity context  and we choose to take into account the simplest one that preserves K\"ahler invariance and amounts to adding a constant uplifting in the scalar potential. Alternatively, we consider the uplifting potential arising from a nilpotent superfield describing the low energy dynamics of a strongly coupled sector. Subsequently,  we study the consequences in inflation using a toy model with one K\"ahler modulus. In Section 4, we study the full model with three K\"ahler moduli and investigate the regions of the parameter space where the scalar potential  is suitable for slow-roll inflation. We derive the scales associated with the inflationary period and  show that our results are in agreement with the Planck 2018 measurements. Finally, Section 5 contains a summary and the  conclusions of our analysis.

 	\section{Review of the geometric stabilisation  mechanism}

 	In a previous paper~\cite{Antoniadis:2018hqy}, working in a type IIB/F-theory framework,  we have considered   a geometric configuration of three intersecting 7-branes with magnetic fluxes in order to solve  the problem of K\"ahler moduli 	stabilisation in a controllable perturbative way. Fluxed 	 7 branes are associated with  anomalous $U(1)$ symmetries which  induce  non-vanishing Fayet-Iliopoulos D-terms and play an instrumental r\^ole in achieving  stabilisation of the volume and K\"ahler moduli. Since D-terms induce positive  contributions to the scalar potential,  they are essential in generating a de-Sitter minimum. Moreover, under certain conditions, to  be described below, the combined effects of the D-terms   and the induced loop-corrections can stabilise all K\"ahler moduli.

 	\subsection{F- and D-term potential}
 	In the presence of D7-branes, the K\"ahler potential receives quantum corrections  which display a logarithmic dependence  on the K\"ahler moduli associated with the size of their transverse dimensions~\cite{Antoniadis:2018hqy}. This  kind of  quantum corrections is a generic phenomenon in string compactifications with branes/localised sources of co-dimenion two and is due to local tadpoles generated by closed strings propagating effectively in two dimensions~\cite{Antoniadis:1998ax}. In type-I string theory for example, logarithmic quantum corrections  have been associated with the volume transverse to D7-branes~\cite{Antoniadis:1998ax}. This effect can  be generalised to anisotropic compactifications of the six-dimensional compact space containing localised fields of co-dimension two.  Based on this fact and using appropriate $T$-dualities~\cite{Antoniadis:2018hqy}  the corresponding corrections in the type IIB/F-theory framework emerge naturally  in the Einstein kinetic term and consequently in the K\"ahler potential defined in the Einstein frame. For the simplest case of one D7 brane, let $u$ be the K\"ahler modulus related to the transverse volume  and $\delta \equiv \gamma \ln u$ the anticipated corrections with $\gamma$ a model dependent parameter. Then, the K\"ahler potential is
 	\be 
 	{\cal K} = -\ln(S-\bar S)-2 \ln(\hat{\cal V}+\xi+\hat\delta)~,
 	\label{Kahler}
 	\ee 
 	where $S$ represents the axion-dilaton modulus and $S-\bar S=2ie^{-\phi}$, with $\phi$ being the 10-dimensional dilaton. Furthermore, 
 	\[ \hat{\cal V}= e^{-3\phi/2}{\cal V},\; \xi= -\frac{\zeta(3)}{4(2\pi)^3g_s^{3/2}}\chi
 	\quad;\quad \hat \delta=\delta g_s^{1/2}\, . 
 	\]
 	In the above formulae, ${\cal V}$ stands for the volume, $g_s=e^{\phi}$ is the string coupling, and  the parameter $\xi$ stands for the ${\alpha'}^3$ curvature perturbative
 	corrections being proportional to the Euler characteristic $\chi$  of the six dimensional compact manifold~\cite{Becker:2002nn}.

 	It has been shown~\cite{Antoniadis:2018hqy} however, that the stabilisation of the K\"ahler moduli requires at least three magnetised 7 branes intersecting each other. 
 	Thus, we consider three K\"ahler moduli $T_1, T_2, T_3$ and define the volume in terms of their real parts $\tau_k=\frac{1}{2}(T_k+\bar T_k)$,
 	\be 
 	{\cal V}= (\tau_1\tau_2\tau_3)^{\frac 12}~\cdot\label{V6}
 	\ee
 	Then,   the K\"ahler potential is
 	\ba
 	{\cal K}&=& -2\ln\left( (\tau_1\tau_2\tau_3)^{\frac 12}+\xi+\sum_{k=1}^3 \gamma_k \ln(\tau_k)\right)~,\label{Kxi123}
 	\ea 
 	where $\gamma_k$ are model dependent coefficients of order one.
 	
 	The cosmological implications of the model and in particular the inflationary properties, to
 	be discussed in the next sections, presuppose the determination of the dynamics of the moduli and other scalar fields.
 	A necessary, although not sufficient, condition that should be fulfilled  at  the vacuum, is the positivity of masses-squared for all moduli
 	fields, with the lighter being the inflaton that we identify with the total six-dimensional internal volume. Since complex structure moduli $z_i$ are assumed to be present in the superpotential, they can be 
 	stabilised by the conditions $D_{z_i}{\cal W}=0$, thus we only need to deal
 	here with the three K\"ahler moduli $\tau_k$. In~\cite{Antoniadis:2018hqy}
 	the minimisation conditions were considered with respect to the total volume ${\cal V}$ and two ratios of the three K\"ahler moduli $\tau_k$. 
 	
 	We first recall the details for the  minimisation with respect to the total volume  of the corresponding F-term scalar potential $V_F$. Assuming for simplicity
 	$\gamma_1=\gamma_2=\gamma_3\equiv \gamma_{\tau} $,   we can
 	write the K\"ahler potential~(\ref{Kxi123})  as follows
 	\be 
 	{\cal K}= -2\log\left({\cal V} + \xi + 2 \gamma_{\tau } \ln({\cal V})\right)~\cdot 
 	\ee 
	The $\alpha'$ corrections, $\xi$, can be absorbed into the logarithmic  term by replacing the implicit string scale unit inside $\ln{\cal V}$ by a new parameter $\mu$
	 \be \xi + 2 \gamma_{\tau } \ln({\cal V}) = 2 \gamma_{\tau } \ln(\mu{\cal V}), \qquad \mu = e^{\frac{\xi}{2\gamma_\tau}} \ee
	  	Computing the corresponding F-term potential, we find
 	\be 
 	V_F= -3\gamma_{\tau}  W_0^2   \frac{\
 		2( \gamma_{\tau}  +2{\cal V}) +(4\gamma_{\tau} -{\cal V})\ln(\mu {\cal V})}
 	{({\cal V}+2 \gamma_{\tau}   \ln(\mu {\cal V}))^2 \left(6 \gamma_{\tau} ^2+{\cal V}^2+8\gamma_{\tau}  {\cal V}+\gamma_{\tau}  (4 \gamma_{\tau}  -{\cal V}) \ln(\mu {\cal V})\right)}~\cdot\label{VF}
 	\ee
 	This  rather complicated formula  obscures
 	the properties of the vacuum of the potential, however, 
 	in the large volume limit and relatively small logarithmic corrections, this can be  approximated by 
 	\be 
 	V_F\approx 3 W_0^2\,\gamma_{\tau} \, \frac{\ln(\mu {\cal V})-4}{{\cal V}^3}~\cdot \label{potV}
 	\ee 
It can be readily inferred that a minimum of the potential (\ref{potV}) exists as long as $\gamma_{\tau }<0$. 
 	Then, imposing   the condition $\frac{d V_F}{d\cal V}=0$  we find that the minimum is localised at $ {\cal V}_{min}=e^{\frac{13}3}/\mu$.
 	Substitution of ${\cal V}_{min}$ into the potential  gives  an AdS minimum
 	\be 
 	{ (V_F)}_{min}= \frac{\gamma_{\tau}  W_0^2 }{{\cal V}^3}<0~,
 	\ee 
 	since $\gamma_{\tau}<0$, as we have already assumed. 	Furthermore, the flatness conditions become
 	 \[D_{\cal V}W_0=-2 \frac{W_0^2 }{{\cal V}_{min}}\ne 0~,\]  
 	 and therefore	supersymmetry is spontaneously broken.

 	The incorporation of $D7$  fluxed branes into the theory, induces also  D-term 	contributions to the effective potential.  
 	Hence, in addition to the F-term potential given above, we take into account the following contribution
 	\ba 
 	V_D&=&\sum_{a=1}^3\frac{d_a}{\tau_a}\left(\frac{ \partial{\cal K}}{\partial \tau_a}\right)^2
 	\approx \sum_{a=1}^3\frac{d_a}{\tau_a^3}
	~,
 	\label{VD}
 	\ea 
	in the large volume limit.
	
 	\noindent
 	Including F- and D-term contributions, in the large volume expansion,  the effective potential can be approximated as follows
 	\[ V_{\rm eff}\approx 3 W_0^2\,\gamma_{\tau} \, \frac{\ln(\mu {\cal V})-4}{{\cal V}^3}+\frac{d_i}{\tau_i^3}+\frac{d_j}{\tau_j^3}+\frac{d_k(\tau_i\tau_j)^3}{{\cal V}^6},\;
 	\; i\ne j\ne k\ne i~\cdot \]
 	The remaining two minimisation conditions determine the ratios between the moduli,  $\left(\frac{\tau_i}{\tau_k}\right)^3=\frac{d_i}{d_k} $. 
 	Expressed in terms of the stabilised total volume ${\cal V}$, these can be written as
 	\[\tau_i^3 = \left(\frac{d_i^2}{d_kd_j}\right)^{\frac 13}{\cal V}^2~\cdot  \]
 	Inserting the above in $V_{\rm eff}$ we obtain the simple form
 	\be 
 	V_{\rm eff}\approx  \gamma \, \frac{\ln(\mu {\cal V})-4}{{\cal V}^3}+\frac{d}{{\cal V}^2}~,
 	\ee 
 	where  for simplicity we have introduced the new constants
 	\ba 
 	d&=&3(d_1d_2d_3)^{\frac 13}~,\label{defd}\\
 	\gamma& =&3W_0^2\gamma_{\tau}~\cdot \label{defgamma} 
 	\ea 
	
It was shown in~\cite{Antoniadis:2018hqy} that the above scalar potential possesses a dS minimum in some region of the parameter space. However since most of the computations were done numerically and the explicit mass spectrum which is necessary for the present study in cosmology was not computed, we perform these computations in the next two subsections below and work out the consequences for inflation.
 	
	\subsection{Normalised fields and mass eigenstates}
Cosmological inflation is conveniently studied  by performing a suitable transformation of the 
inflaton field in order to obtain  a canonically normalised kinetic term. 
Obviously, this is also needed for the computation of the mass spectrum.

\noindent
We start by  first noting that in the limit where the volume 
$\V$ is much larger than the loop correction term $\gamma_i \LN (\frac{\V}{\tau_i})$, we can use the no-scale K\"ahler potential  $\K_{NS}$
\be
\K_{NS} = -2\LN (\V) = -\sum_i \LN (T_i + \ov{T}_i) = -\sum_i \LN (\tau_i)~,
\ee
to derive the kinetic terms. We obtain
\ba
\K _{T_I\ov{T}_J}
\partial T_I \partial \ov{T}_J  = \sum_i\frac{\partial T_i \partial \ov{T}_i}{(T_i + \ov{T}_i)^2} = \sum_i\frac{(\partial \tau_i)^2}{4\tau_i^2} + \cdots  = \sum_i\frac{1}{2}(\partial t_i)^2 + \cdots , 
\ea
where $\{\cdots\}$ correspond to the kinetic terms of axions  absorbed by the $U(1)$ fields related to the magnetised branes with the corresponding D-terms, and
\be 
t_i = \frac{1}{\sqrt{2}}\LN (\tau_i)~,
\ee
are the normalised real scalar fields. Since we have three K\"ahler moduli, it would be better to switch to the mass eigenstates 
and find the lightest one. One option is taking the total volume (\ref{V6}) and two other perpendicular directions:
\ba
t &&= \frac{1}{\sqrt{3}}(t_1 + t_2 + t_3) = \frac{\sqrt{6}}{3} \LN (\V)~,\label{ttoV}\\
u &&=  \frac{1}{\sqrt{2}} (t_1 - t_2)~,\\
v &&=  \frac{1}{\sqrt{6}} (t_1 + t_2 - 2t_3)~\cdot 
\ea
The corresponding F-term  and D-term components of the scalar potential in this new basis are
\ba
V_F &&\simeq \frac{\gamma}{2} e^{-\frac{3\sqrt{6}}{2}t} (\sqrt{6}t + 2\ln (\mu) - 8) 
~,\label{VFt}\\
V_D &&\simeq \frac{d_1}{\tau_1^3} + \frac{d_2}{\tau_2^3} + \frac{d_3}{\tau_3^3} 
\\
&&= d_1 e^{-\sqrt{6} t - \sqrt{3} v - 3u} + d_2 e^{-\sqrt{6} t - \sqrt{3} v + 3u} + d_3 e^{-\sqrt{6} t + 2\sqrt{3} v} 
~,
\label{VDtuv}
\ea
where the $\simeq$ sign is used for the  volume suppressed corrections. 

A deviation from the assumed universal value $\gamma_i=\gamma_\tau$ contributes an additional term linear in $t_i$ inside the bracket of eq. (\ref{VFt}). In the $t$, $u$, $v$ basis, the additional part of the potential can be written as:
\be \delta V_F \simeq  e^{-\frac{3\sqrt{6}}{2}t} (\delta\gamma_t t + \delta\gamma_u u + \delta\gamma_v v). \label{NUVF}\ee
As long as $\delta\gamma_u$ and $\delta\gamma_v$ are smaller than the universal $\gamma$, the $u$ and $v$ dependent parts in (\ref{NUVF}) are negligible compared to the D-term potential (\ref{VDtuv}), and we ignore them. Then, together with $\ln (\mu)$, eq. (\ref{VFt}) can be written as
\bea
V_F &=& \frac{\gamma}{2} e^{-\frac{3\sqrt{6}}{2}t} (At + B),\\
&\sim& \frac{\gamma}{2} e^{-\frac{3\sqrt{6}}{2}t} (\sqrt{6}t - 8 + 2x)\label{rescaleVF}\eea
where $A$ and $B$ are $t$-independent constants. The non-universal parts contribute to $A$ and $B$ while $\ln (\mu)$ contributes   to  $B$ only. Since the overall scale is irrelevant in slow-roll calculation, we can rescale $A$ to be the same as in eq. (\ref{VFt}) and the parameter $ x$ in eq. (\ref{rescaleVF}) contains both non-universal terms and $\ln (\mu)$. In the next subsection, we will see that the existence
of a non-vanishing $x$ will only shift the potential towards larger t. Furthermore, the calculation of the slow-roll parameters 
 are dominated by the exponential part $e^{-\frac{3\sqrt{6}}{2}t}$. Thus, for simplicity we neglect both non-universal terms 
 and $\ln (\mu)$  in eq. (\ref{VFt}).
  
The vanishing of the first derivatives with respect to 
the moduli $u$ and $v$ of (\ref{VDtuv}) lead to the minimisation conditions
\be
\begin{split}
	u_0&=\frac{1}{6} \ln \left(\frac{d_1}{d_2}\right)\\
	v_0&=\frac{1}{6 \sqrt{3}}\ln \left(\frac{d_1 d_2}{d_3^2}\right)~.
\end{split}
\ee
At these positions, the D-term potential  becomes
\be 
V_D|_{u_0, v_0} = e^{-\sqrt{6} t} 3(d_1d_2d_3)^{\frac{1}{3}} =d\, e^{-\sqrt{6} t}~,
\ee
where $d$ is defined in~(\ref{defd}).

\noindent
The masses of $t$, $u$ and $v$ can be calculated by taking the second derivatives of equations (\ref{VFt}) and (\ref{VDtuv})
\be
\begin{split}
m^2_t& = V_{Ftt} + V_{Dtt}|_{u_0, v_0} = \frac{9}{4}\gamma e^{-\frac{3\sqrt{6}}{2}t} (3\sqrt{6} t - 28) + 6d\, e^{-\sqrt{6} t}~,\\
m^2_u &= V_{Duu}|_{u_0, v_0} = m^2_v = V_{Dvv}|_{u_0, v_0} =6 d\,e^{-\sqrt{6} t}~,
\end{split}
\ee
with $\gamma $ defined in~(\ref{defgamma}).

The question now arises whether the potential meets the requirements
for a slow-roll single field inflation where the inflaton field is played by some K\"ahler 
modulus. It is readily realised that the most natural candidate is the total volume.
Hence, in the regime where the moduli masses satisfy the condition
\beq 
|m_t^2| \ll m_u^2 = m_v^2~,\label{masscondition}
\eeq
we can use the effective potential
\ba
V_{\rm eff}= 	V_F + V_D|_{u_0, v_0} &=& \frac{1}{2}\gamma \, e^{-\frac{3\sqrt{6}}{2}t} (\sqrt{6}t - 8)+ d\, e^{-\sqrt{6} t} ~,
\label{Vinf0}
\ea
to describe the dynamics of the modulus $t$ related to the total volume via~(\ref{ttoV}).
\subsection{A first step towards inflation}

To examine the implications on inflation, firstly we proceed with an analysis of the properties of the scalar potential. 	
At the classical level, we already know that the K\"ahler moduli dependent scalar potential vanishes identically. Then, 
the $t$-dependent  potential $V_{\rm eff}$ in eq.(\ref{Vinf0}) generated from quantum corrections, in the absence of 
any other effects -including the FI-terms to be discussed later- should vanish as $t\to \infty$.
Moreover, the requirement for a dS minimum implies that its zero value at infinity should be approached from positive values 
$\underset{t\to\infty}{\lim}V_{\rm eff}(t)\to 0^+$. In the simplest case, this can be realised when the potential has a minimum and a maximum at finite
values of the volume modulus $t$.

\noindent
We start the analysis  with the minimisation of the rescaled potential in eq. (\ref{rescaleVF})
\be
V_{\rm eff}=  \frac{1}{2}\gamma \, e^{-\frac{3\sqrt{6}}{2}t} (\sqrt{6}t +2x - 8)+ d\, e^{-\sqrt{6} t} ~,
\label{Veffx}
\ee
with respect to $t$. In searching for extrema of the potential, 
we put its first derivative equal to zero and  obtain  the equation
\ba                                                                         
V_{\rm eff}'  =   \frac{d V_{\rm eff}}{dt} = -\frac{\sqrt{6}}{4} e^{- \frac{3\sqrt{6}}{2} t} \left( 4 d e^{\frac{\sqrt{6}}{2} t} +
\gamma  (3\sqrt{6}t + 6x -26)\right) = 0~\cdot\label{Veffx'}
\ea
This equation has two solutions which correspond to a minimum and a maximum of the potential. 
Indeed, defining 
\be
w=\frac{13}3 - x - \frac{\sqrt{6}}{2}t\label{wVol}~,
\ee  
the above condition takes the form
\be 
w\, e^w =z,\;\; {\rm with}\;\; z= \frac{2}{3}\frac{d}{\gamma}e^{\frac{13}{3}-x}
~.
\label{mincondV0}
\ee 
For real values, $w$ should be restricted by the lower bound $w>-e^{-1}$. Moreover, the  solution 
of (\ref{wVol}) is double-valued in the 
range $(-e^{-1},0)$, and it is expressed in terms of the   Lambert $W$-function. 
This range in turn implies the following bounds on the ratio of the parameters $d$ and $\gamma$  
\beq 
-\frac{3 e^{-\frac{16}{3}}}{2} \simeq - 0.007242 < \frac{d}{\gamma} e^{-x}< 0~.\label{2extremecondition}
\eeq

\noindent 
The  two branches of the Lambert $W$-function are denoted with $W_0(z), W_{-1}(z)$ and determine the minimum and the maximum  of the 
canonically normalised `volume'  respectively,  through the equations
\ba
t_{min} &=&\sqrt{\frac{2}{3}} \left(\frac{13}3 - x - W_0\left(\frac{2de^{\frac{13}{3} - x}}{3\gamma}\right)\right),\label{tmin}\\
t_{max} &=& \sqrt{\frac{2}{3}} \left(\frac{13}3 - x - W_{-1}\left(\frac{2de^{\frac{13}{3} - x}}{3\gamma}\right)\right).\label{tmax}
\ea

\begin{figure}[h!]
	\centering
		\includegraphics[width=0.65\columnwidth]{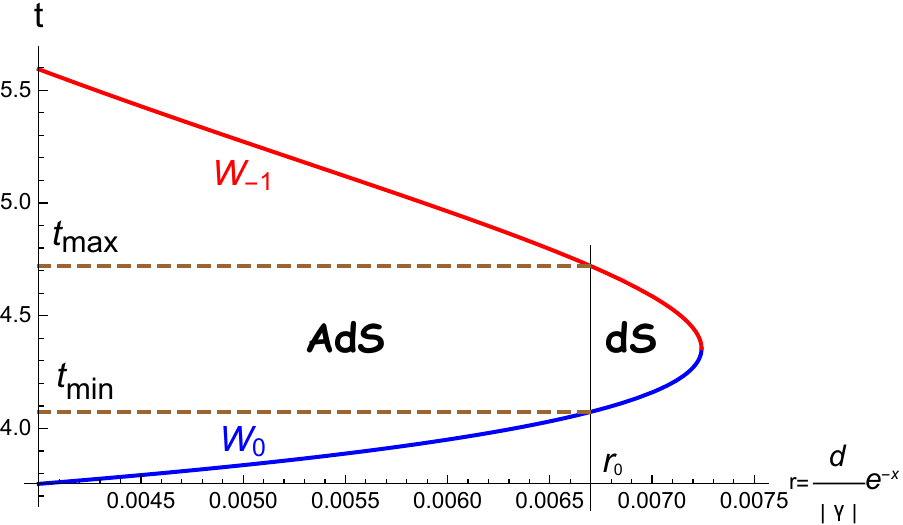}
	\caption{
		\footnotesize
		{The canonically normalised volume plotted as a function of the parameter ratio $r=\frac{d}{|\gamma|} e^{-x}$.
		The two different colours of the curve correspond to the two different branches of the Lambert function(see text). 
		For this case	de Sitter minima are only possible for $r\ge r_{0}$.  	
		The positions of the minimum and maximum of the scalar potential are 
			denoted with $t_{min},t_{max}$ respectively. }
	}
	\label{VinfN=60old}
\end{figure}

\noindent 
The region~(\ref{2extremecondition}) is further restricted from above, 
due to the requirement for a positive energy minimum of the scalar potential. Putting eq.~(\ref{tmin}) into eq. (\ref{Veffx}), one gets
\be V^{\rm eff}_{min} = \frac{1}{3} e^{-\frac{3\sqrt{6}}{2}t_{min}} \left(\gamma + d e^{\frac{\sqrt{6}}{2}}t_{min} \right) \ge 0,\ee
which can be solved to give the numerical 	bound  $\frac{d}{\gamma}e^{-x} \le -0.006738$. Hence, finally the range is restricted to be 
\beq 
- 0.007242 < \frac{d}{\gamma} e^{-x}\le  -0.006738~,\label{2extremecondition2}
\eeq
as can be seen in Fig.~\ref{VinfN=60old}.

\noindent 
Given the fact that the form of the potential~(\ref{Veffx}) ensures a positive vacuum energy, in the following  we would like to  examine whether  it is also sufficient to accommodate inflation.
We will see that this is not possible and new contributions have to be taken
into account, which will be discussed in the subsequent section. 

Indeed, the region~(\ref{2extremecondition2}) is too restrictive to accommodate the  number of e-foldings required in the slow roll inflationary
scenario. To show this, it suffices to work out  the limiting case  $\frac{d}{|\gamma|}e^{-x} = r_0 \approx 0.006738$,  corresponding to the upper bound 
of (\ref{2extremecondition2}) where the two extrema  of the effective potential   are maximally separated. 
In this case the minimum	and the maximum of the potential are found to be
\[  t_{min}\approx 4.07 - \sqrt{\frac{2}{3}}x, \;\; t_{max}\approx 4.72 - \sqrt{\frac{2}{3}}x
\]
The ratio between the potential $V_{\rm eff}$ (\ref{Veffx}) and its derivative 
$V'_{\rm eff}$ (\ref{Veffx'})
\be \frac{V_{\rm eff}}{V'_{\rm eff}} = -\frac{1}{\sqrt{6}} + \frac{3 (t + \sqrt{\frac{2}{3}}x) - 5\sqrt{6}}{3 \left( 3\sqrt{6} (t + \sqrt{\frac{2}{3}}x) + 4 r e^{ \sqrt{\frac{3}{2}}(t + \sqrt{\frac{2}{3}}x)} -26\right)}
\ee
is always a function of $(t + \sqrt{\frac{2}{3}}x)$.
Using this formula, we find that  
the number of e-foldings for this limiting case are within the bound
\be 
N \leq\int_{t_{min}}^{t_{max}} \frac{V_{\rm eff}}{V'_{\rm eff}}\,dt\approx {\cal O}(1)~,
\ee 
which is independent of $x$. 

\noindent 
The number of e-foldings is an order of magnitude smaller than required.
The  reason is that in a large part of the parameter space $\frac{d}{\gamma}\in [-0.006738$-$0.0]$ where 
the difference $\Delta t= t_{max}-t_{min}$ would be sufficient to 
give the observed number of e-foldings, the potential displays an AdS minimum. 
Therefore, we infer that  it is impossible to generate enough e-foldings without 
a new contribution which will uplift the potential.

 	\section{Additional Uplifting}
 	 \subsection{Contributions from Fayet-Iliopoulos terms}
 	 
From the discussion in the previous section, we have realised that 
  the existence of an uplifting term  is of crucial importance to  implement  the inflationary scenario in the present model. Constant Fayet-Iliopoulos D-terms would be a possible tool to use in this direction. However, unlike in global supersymmetry, such terms are highly restricted in supergravity and even more in string theory. Part of the problem is due to the fact that at the Lagrangian level a constant FI-term is not gauge invariant. Moreover, a constant FI-term in supergravity can be written only for a $U(1)$ that gauges the $R$-symmetry and thus a constant term in the superpotential is not allowed. In the absence of matter this leads  to the Friedman supergravity model~\cite{Freedman:2012zz} that breaks supersymmetry in de Sitter (dS) space without superpotential and, thus, without explicit gravitino mass. It is unclear how this model (or generalisations of it) could be realised in string theory.
 	
 	Recently, Ref.~\cite{Cribiori:2017laj} has initiated a new class of FI-terms in supergravity that avoid the above problems. Indeed, a novel FI-term was proposed which is gauge invariant at the Lagrangian level and can be written for a non-R $U(1)$. Although its form appears to be non-local in superspace, it can be expanded in components if the D-auxiliary field has non-vanishing expectation value (VEV), leading to the standard FI-term of global supersymmetry plus fermionic contributions involving the $U(1)$ gaugino. In the case where supersymmetry is broken only by this D-term, the gaugino is absorbed by the gravitino that becomes massive due to the super-Higgs mechanism and the Lagrangian is reduced to the usual FI-term. In the absence of matter, one obtains the Friedman model extended by a constant superpotential that amounts to a gravitino mass-term $m_{3/2}$ as independent parameter from the cosmological constant that shifts the vacuum energy above the anti-de Sitter lower bound of $-3m_{3/2}^2$. In the presence of neutral matter, the new D-term leads to a scaler potential $\sim e^{2{\cal K}/3}$, with ${\cal K}$ the K\"ahler potential, that breaks K\"ahler invariance of standard $N=1$ supergravity explicitly. Interestingly, for the case of the total volume, this potential is similar to the one of $\overline{D3}$ brane.
 	
 	The new FI-term has been extended in several ways. In particular, it has been shown that it can be written in the presence of charged matter superfields and on top of the standard FI-term, even in the case of R-symmetry~\cite{Antoniadis:2018cpq}. Moreover, it is possible to consistently modify it so that K\"ahler invariance is preserved~\cite{Antoniadis:2018oeh}, in which case it is constant and the vacuum energy is uplifted by an arbitrary constant value. On the other hand, new functions of chiral matter field may be involved in more general extensions of this term~\cite{Farakos:2018sgq, Aldabergenov:2018nzd, Antoniadis:2018cpq}.
 	
 	It is an open interesting question wether such terms can appear in the effective supergravity of string compactifications. Here we will consider the consequences of adding a constant FI-term as an uplifting vacuum energy source in the effective supergravity that allows K\"ahler moduli stabilisation due to perturbative string corrections, discussed above. We will assume that this FI-term is associated with an extra $U(1)$ whose gauge coupling is fixed by 3-form fluxes, such as in the case of an effective 3-brane. Denoting this constant contribution with $V_{up}$ and taking into account the F- and D- contributions discussed above, the final form of the scalar potential becomes:
 		\be 
 	V_{\rm eff}\approx W_0^2\,\gamma_{\tau} \, \frac{\ln({\cal V})-4}{{\cal V}^3}+\frac{d}{{\cal V}^2}+V_{up}\label{VFDVup}~.
 	\ee
	
	 	\subsection{Contribution from a nilpotent supefield}
Alternatively, one could consider the uplifting potential from strongly coupled matter fields \cite{Dudas:2006gr,Abe:2006xp}. In our scenario, we expect these matter fields to be localised at the intersection points of the three 7-branes.  After integrating out the heavy modes, the effective K\"ahler potential and superpotential contain an additional nilpotent superfield $X$ with $X^2 = 0$. \footnote{The nilpotency condition arises as an effective description of the models of \cite{Dudas:2006gr,Abe:2006xp} after integrating out the scalar component in $X$.}
\ba \Delta \mathcal{K} &=& \frac{X\ov{X}}{\tau_1^{n_1} \tau_2^{n_2} \tau_3^{n_3}},\\
\Delta \mathcal{W} &=& f X,\ea
 where $f$ is a dimension 2 constant related to the dynamical supersymmetry breaking sector,
 and  $n_i$ are rational numbers correspondiing to modular weights of the dual heterotic theory. In orbifold compactifications they are 
 given by a sum of integers (number of oscillators) and orbifold twists~(see for instance~\cite{Ibanezbook}).

 \noindent 
  We assume the universal modular weights $n_1 = n_2 = n_3 = n$ leading to the additional F-term potential in the large volume limit is
 \be 
 \Delta V_F = \frac{f^2}{(\tau_1\tau_2\tau_3)^{1-n}} = \frac{f^2}{\mathcal{V}^{2-2n}}.
 \ee
 
  \noindent 
 Here we consider the interesting case where $n=1 - \frac{1}{2} \nu$ with $\nu$ being an orbifold twist with possible values in supersymmetric orbifolds
 $\nu=0$ or multiples of $\frac{1}{12}$. 
 If we take $\nu = 0$, this shares the same form as the uplifting term from the new D-term in eq. (\ref{VFDVup}) with $V_{up} = f^2$.
 
 We have checked that $\nu=\frac{1}{12}$ provides  a realistic model of inflation. However, below we present the analysis for the simplest 
 case $\nu=0$ which reproduces also the contribution on the new FI term in~(\ref{VFDVup}).  Thus, below we will use the effective potential
 	\ba
 	V_F + V_D|_{u_0, v_0} + V_{up}&=&  \frac{1}{2} e^{-\frac{3\sqrt{6}}{2}t} (\sqrt{6}t - 8)\gamma + e^{-\sqrt{6} t} d + V_{up}~,\label{Vinf}
 	\ea
 	to describe the dynamics of the modulus $t$ related to the total volume via~(\ref{ttoV}), where $\gamma$ is defined in eq. (\ref{defgamma}).

  \subsection{A toy model: One K\"ahler modulus inflation}
\begin{figure}[H]
 		\centering
 		\includegraphics[width=0.80\columnwidth]{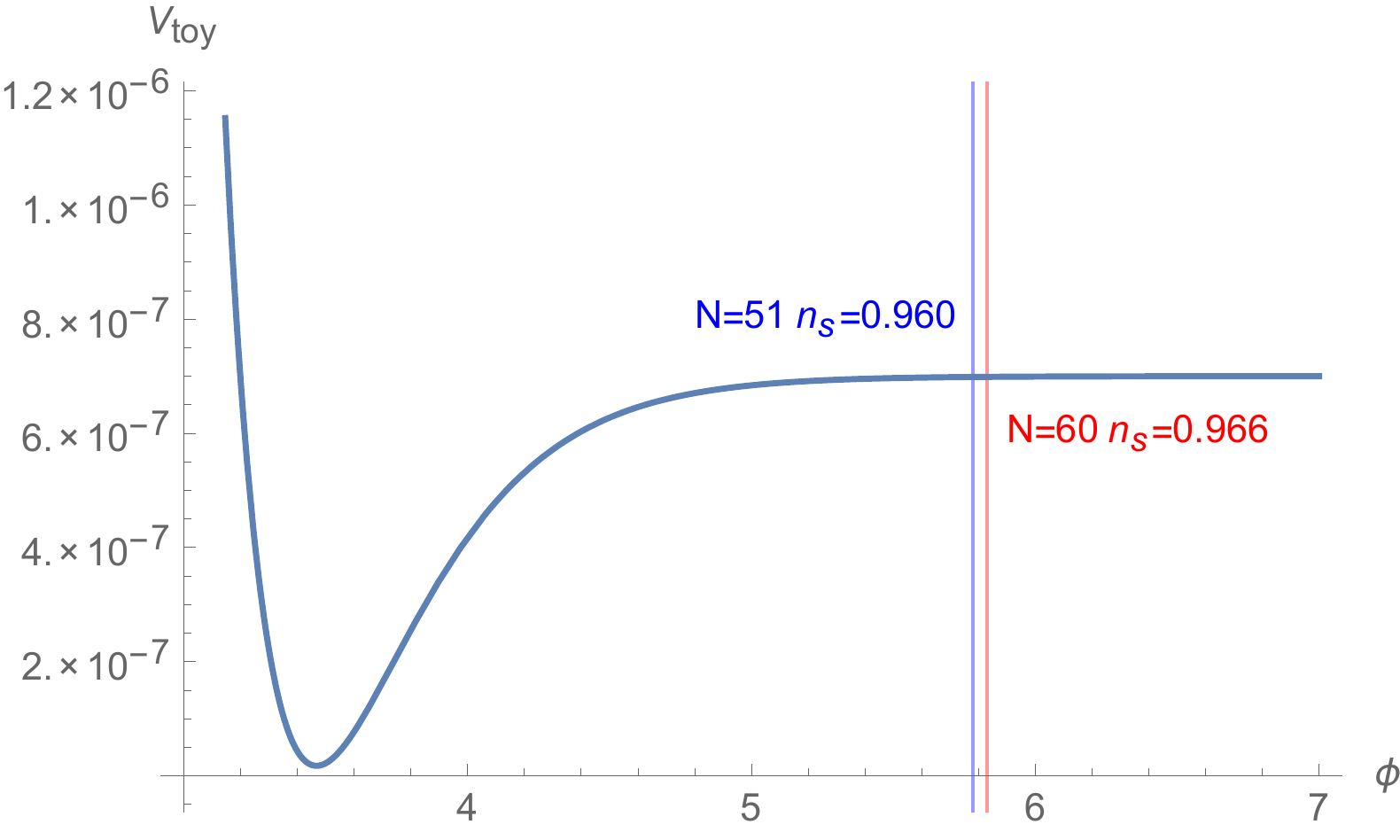}
 		\caption{
 			\footnotesize
 			{The effective potential of eq. (\ref{1KMpotential}) with $\gamma = -0.7, \mathcal{W}_0 = 1, V_{up} = 7\times 10^{-7}$.}
 		}
 		\label{1KMtoyfig}
 	\end{figure}

 	Before analysing the fully-fledged case,  we can first consider a toy model which contains only one K\"ahler  modulus,
 	 and turn on the logarithmic correction and the cosmological constant. The corresponding K\"ahler potential and the effective potential in the large volume limit reads
	\ba  	{\cal K}_{toy} &&= -2 \ln(\tau^{\frac{3}{2}} + \gamma_{toy} \ln(\tau))~,\\
	V_{toy} &&\simeq  \gamma_{toy}\mathcal{W}_0^2\frac{ 3 \ln (\tau) - 8}{2\tau^{\frac{9}{2}}} 
 +V_{up}\\
	 &&=  \gamma_{toy}\mathcal{W}_0^2\frac{ \sqrt{6} \phi - 8}{2} e^{-\sqrt{\frac{27}{2}}\phi} 
	+ V_{up}~,\label{1KMpotential}
	\ea
 	where we use in the last line the canonically normalised field $\phi = \sqrt{\frac{3}{2}} \ln (\tau)$. Eq. (\ref{1KMpotential}) gives a form similar to Starobinsky-like inflation and the slow-roll parameters fit well with the observations as shown in Fig.~(\ref{1KMtoyfig}).
 	However, for the realistic case, as discussed in \cite{Antoniadis:2018hqy}, one needs at least three K\"ahler moduli to induce the logarithmic corrections. Thus we need also to stabilise the other two moduli during inflation.

 	\section{Inflation from the total volume}

  In the following, we analyse the general case including corrections to the effective potential due to all FI-terms.

 	 Inflation should occur between the maximum $t_{max}$ and the minimum, $t_{min}$ of eqs.~(\ref{tmin}-\ref{tmax}). 
 	 Between them, we can define another two points. The first one is the ending point of inflation, which corresponds 
 	 to the breaking of the slow-roll condition:
 	\beq
 	t_{end} ={\rm max}\{t |_{\frac{1}{2}\left(\frac{V'}{V}\right)^2 \simeq 1}, t |_{\frac{|V''|}{V} \simeq 1}\}, 
 	\eeq
 	where the derivatives $V', V''$ are taken with respect to $t$. The second point is the one corresponding to the 
 	anticipated number of e-foldings
 	 $N_*\sim 50$ to 60, where $N_*$ is given by the formula
 	\beq
 	 N_* = \int_{t_{end}}^{t_{*}} \frac{V}{V'}dt~\cdot
 	 \eeq
 	In addition, at the same point $t_* (N_*)$, the spectral index should satisfy the values from observations
 	\beq 
 	n_{s} = 1-6\epsilon+2\eta=1 - 3\left.\left(\frac{V'}{V}\right)^2\right|_{t_*} + 2\left.\frac{V''}{V}\right|_{t_*}~\cdot\label{spectralindex}
 	\eeq
Obviously, the four positions should satisfy the inequality:
 	\beq
 	t_{min} < t_{end} < t_* \leq t_{max}.
 	\eeq
 	Due to the presence of the other two  scalar fields, $u$ and $v$, we should consider the multi-fields effect which happens when the inflaton is no longer the lightest scalar, and  the condition eq. (\ref{masscondition}) is violated:
 	\be 
 	t_{mf} \in \{t | {|m_t^2| \geq m_{u/v}^2}\}~\cdot 
 	\ee
 	
\begin{figure}[H]
 		\centering
 		\includegraphics[width=0.80\columnwidth]{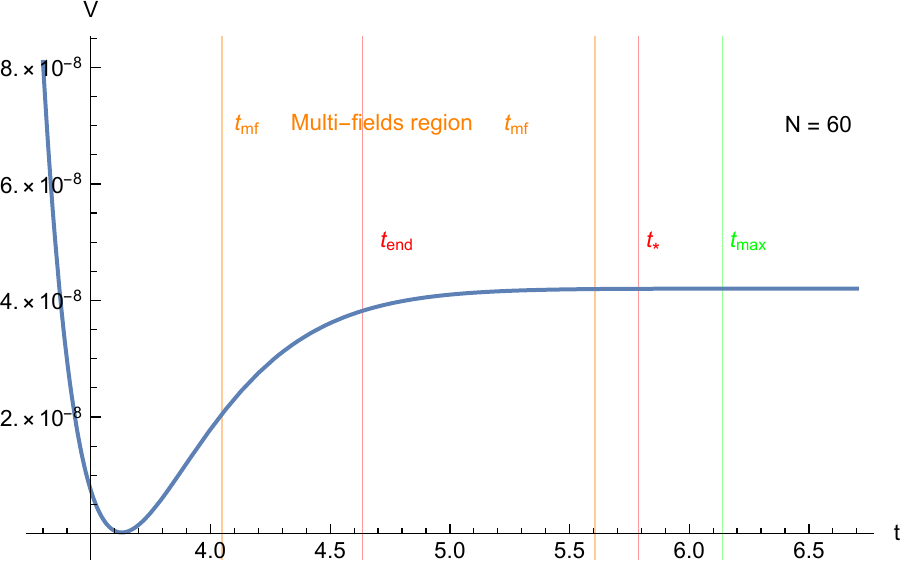}
 		\caption{
 			\footnotesize
 			{The potential (\ref{Vinf}) with the parameters from Table~\ref{parameterstable} for $n_s = 0.9605, N_* = 60$}
 		}
 		\label{VinfN=60}
 	\end{figure}

\begin{figure}[H]
 		\centering
 		\includegraphics[width=0.80\columnwidth]{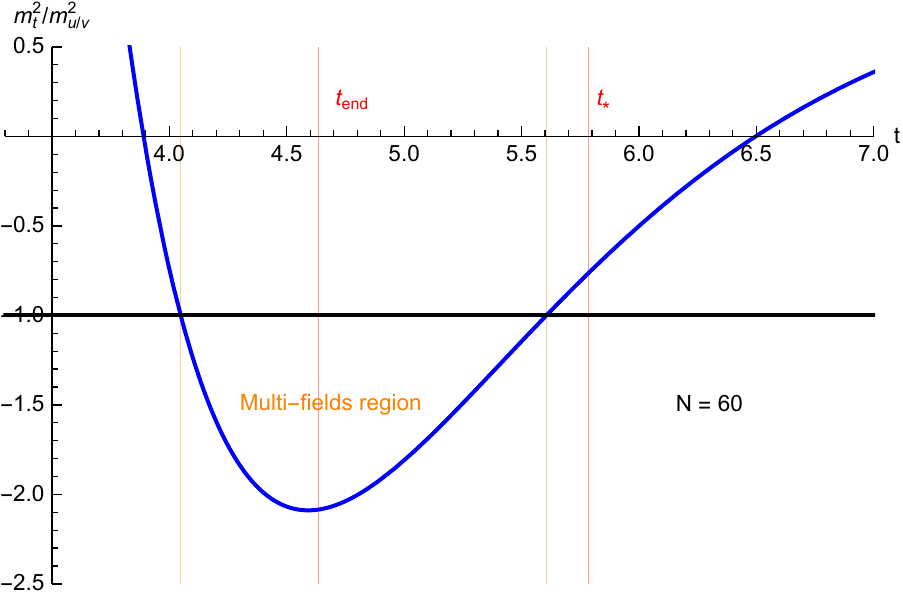}
 		\caption{
 			\footnotesize
 			{The mass ratio between the inflaton $t$ and the other two K\"ahler moduli for  $n_s = 0.9605, N_* = 60$. The multi-field region is among the two orange lines.}
 		}
 		\label{massratio}
 	\end{figure}

\begin{figure}[H]
 		\centering
 		\includegraphics[width=0.80\columnwidth]{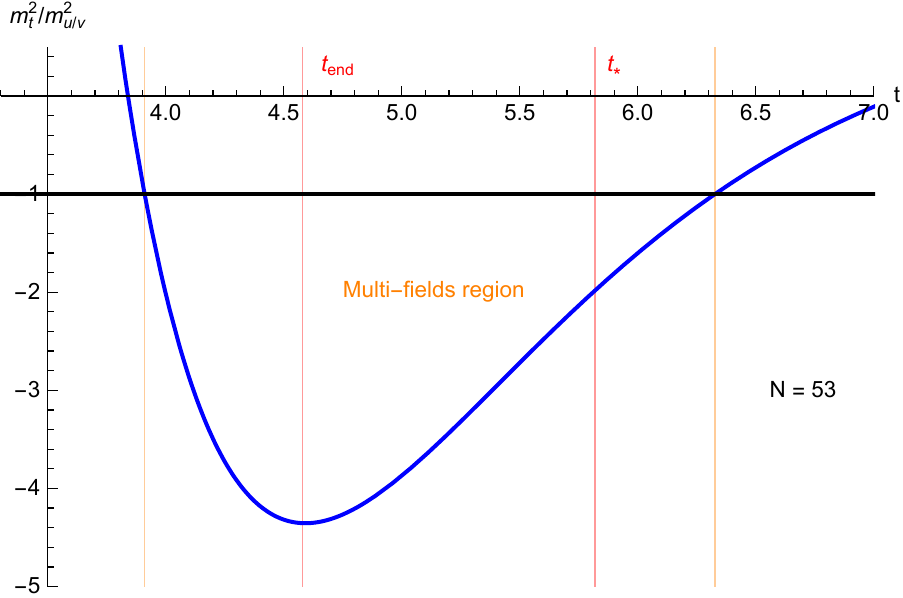}
 		\caption{
 			\footnotesize
 			{The mass ratio between the inflaton $t$ and other two K\"ahler modulus for  $n_s = 0.9605, N_* = 53$. The multi-field region is among the two orange lines which covers the whole inflation period.}
 		}
 		\label{massratioN=53}
 	\end{figure}

 	\noindent 
	We now study the parameter space to find regions satisfying all the aforementioned constraints. In the large volume limit, the inflationary potential~(\ref{Vinf}) has three parameters. There are three conditions to fix each one of them: the first is that at the minimum, the vacuum energy is almost zero; the second is that at the point $t_* (N_*)$ corresponding to a given number $N_*$ of e-foldings, the spectral index $n_s$ should fit the observed value; the last one is the scalar power spectrum amplitude $A_s$. Since the three parameters are all linear in the large volume limit, one can first absorb the overall scale and fix $\gamma$ to be a small negative number. In fact, the choice of $\gamma$, although redundant, is convenient for the numerical analysis. Thus, taking $\gamma=-0.1$, the scalar potential can be written as
 	\ba
 	 V_{inf} &=& V_F + V_D|_{u_0, v_0} + V_{up}\nonumber\\
 	 & =&  a\,(-\frac{1}{20} e^{-\frac{3\sqrt{6}}{2}t} (\sqrt{6}t - 8) + e^{-\sqrt{6} t} d + V_{up}) \label{Vinfrepa}~,
 	\ea
 	in which $a$ is the overall constant which can be fixed by the amplitude later. The almost zero vacuum energy in the minimum can fix another parameter. We choose $V_{up}$ to be the one fixed and write it as a function of $d$,  $V_{up}=V_{up}(d)$. Finally, the spectral index $n_s$ observed fixes the last parameter, in terms of the e-fold number $N_*$: $d(n_s, N_*)$. Numerically we solve the condition eq. (\ref{spectralindex}) and find that $d(N_*)$ is a monotonically increasing function in the range $50 \leq N_* \leq 60$ for  $n_s = 0.9605$ which is within the  $1~\sigma$ region of Planck 2018 TT,TE,EE+lowE measurement \cite{Akrami:2018odb}. The results are shown in Table~\ref{parameterstable}.
 	\begin {table}[H]
 	\begin{center}
 		\begin{tabular}{ | c | c | c | c | c | c | c | c | c | c | c |}		
 			\hline	
 			$N_*$ & $\gamma$ & d & $V_{up}$ & $a$ & $r$ & $t_{min}$ & $t_{end}$ & $t_*$ & $t_{max}$ & $t_{mf}$  \\
 			\hline
 			53 & -0.1 & 0.00015 & $5.85 \times 10^{-8}$ & 0.000111 & 0.00021 & 3.57 & 4.58 & 5.82 & 6.76 & $3.91 \leq t \leq 6.33$\\
 			\hline
 			60 & -0.1 & 0.00026 & $4.20 \times 10^{-8}$ & 0.000074 & 0.00010 & 3.63 & 4.63 & 5.79 & 6.14 & $4.05 \leq t \leq 5.61$\\
 			\hline
 		\end{tabular}
 		\caption{\footnotesize{The predictions for two choices of the parameters $\gamma, d, V_{up}, \alpha$ involved in the scalar 
 				potential~(\ref{Vinfrepa}), for fixed value of the spectral index $n_s = 0.9605$.}}\label{parameterstable}
 	\end{center}
 	\end {table}
 	We show one example for $N_* = 60$ in Fig.~\ref{VinfN=60}.
 	Between the two orange lines is the area that the inflaton $t$ is not the lightest K\"ahler modulus. This can be seen from the ratio of the different mass scales in Fig.~\ref{massratio}.
 	As we decrease $N_*$ by lowering the parameter  $d$, we can see that the multi-field region becomes larger. In the example of $N_* = 53$ in Table~\ref{parameterstable}, the multi-field region covers the whole inflation period as shown in Fig.~\ref{massratioN=53}.
 	
	The existence of a multi-field region brings the question of possible large non-gaussianities. Here,
	in both cases, we assume that the moduli ratios  are stabilised at the minimum during the whole period of inflation. There could be a deviation from the minimum at the initial time $t_*$. In the case of $N_*=60$ above, since inflaton is the lightest scalar at $t^*$, one can set the initial condition of deviation to be small. The mass of the ratios are much below the Hubble expansion rate $H$, so that the Hubble damping will take them back to their minimum without oscillations. The fraction of the kinetic energy of the moduli ratios thus keeps decreasing. From \cite{Byrnes:2008wi}, this leads to negligible multi-field effects in the non-gaussianity signals. One can treat them as effective single field inflation. In the case of $N_*=53$, the trajectory we choose is still effectively single field inflation. However the initial condition is more fine-tuned since the inflaton at $t^*$ is already heavier than the ratio moduli.

 	\section{Discussion and Conclusions}

 	In the present work,  slow roll inflation has been studied in the framework of type II-B/F theory effective supergravity
 	using the six-dimensional compactification volume modulus as the inflaton.  The analysis is performed in the context of 
 	a recently proposed framework of moduli stabilisation~\cite{Antoniadis:2018hqy}, in which  a geometric configuration of three intersecting stacks 
 	of D7 branes is considered. In  this set up, perturbative string loop  contributions  induce terms in the K\"ahler potential
 	which depend logarithmically on the volume moduli  associated with the directions transverse to the corresponding D7 branes. These  contributions,
 	together with the	-already well known- $\alpha'$ corrections, break the no-scale invariance of the K\"ahler potential.
 	Also, the aforementioned D7 branes with magnetic fluxes induce positive definite D-terms in the  scalar potential. These 
 	two elements, the logarithmic loop-corrections and the positive D-terms, 
 	are sufficient to stabilise the K\"ahler moduli. At the same time, they ensure  a de Sitter minimum with a tiny positive 
 	cosmological constant.  It is worth emphasising that this geometric stabilisation method, does not require inclusion 
 	of  non-perturbative  effects in the superpotential. 
 	The  dS vacua generated with the use  of the above ingredients only, are restricted in a small region of the available parameter space. 
 	Then, regarding the implementation of a successful cosmological inflationary scenario, it can be easily realised that this is not possible because the predicted number of e-foldings  $N_*$
 	 are found to be around an order of magnitude smaller compared to the anticipated value $N_*\approx 60$. 
 	 
 	 \noindent 
As a result, it seems that within our perturbative large volume approximation, one needs a new source of uplifting the vacuum energy.
In this work we considered the effect of a new FI contribution to the scalar potential written recently in global and local supersymmetry which is manifestly gauge invariant at the Lagrangian level and, thus, avoiding the requirement of standard supergravity to be associated to a $U(1)$ R-symmetry~\cite{Cribiori:2017laj, Antoniadis:2018cpq, Antoniadis:2018oeh}. This opens also the possibility of being generated as an effective term in string compactifications, although this is not clear at present.
 	Alternatively, we considered a class of uplifting potentials arising from a nilpotent superfield and depending on some rational modular weights. A particular choice reproduces the effect of the new FI term. Although realistic inflation can be obtained in several cases, here we presented a detailed analysis for the simplest case that describes also the new FI term. When the latter is included   in the scalar potential, new metastable de Sitter vacua can be generated which are suitable 
 	for slow-roll cosmological inflation.
 	More precisely, the effective potential derived in this geometric set-up depends on  three unspecified 
 	parameters  multiplying an equal number of simple distinct elements:  
 	the first two are the strentghs of the F-term (which includes the $\alpha'$ and the logarithmic string loop corrections)  and the D-term,
 	and the third parameter is associated with the new FI-term. Moreover, F- and D-terms depend on the three  K\"ahler moduli related to
 	the three intersecting D7 branes.

 	\noindent 
 	The inflaton field is identified  with the logarithm of the internal $6d$-volume,  and it is
 	proportional to the sum  of the three canonically normalised K\"ahler moduli.
 	This choice is suggestive for a new basis, in which the remaining two degrees of freedom (combinations of the K\"ahler moduli)
 	are taken to be `orthogonal' to   the total volume. F- and D-terms are  instrumental in defining the mass spectrum of the three moduli fields via 
 	the minimisation procedure, while the new FI-term ensures a sufficiently positive vacuum energy to accommodate slow-roll inflation. 
 	After implementing the constraints imposed by the minimisation conditions, the scalar potential obtains a simple form. In the large volume limit
 	all slow roll conditions are satisfied  by fitting  the three available parameters.   	
	The model predicts  a relatively small value
 	for the tensor-to-scalar ratio parameter ($r\sim 0.0002$) which is far beyond the required values for observation. 
 	Although a definite  clue that inflation has a string theory  origin  is still lacking,  the successful implementation
 	of the slow-roll inflation in the present model   provides  a very compelling argument for the relevance of string theory  to cosmology. 
 	
 	\vspace*{1cm}

  	 \section*{Acknowledgements}
  	This work was supported in part by the Swiss National Science Foundation, in part by Labex ``Institut Lagrange de Paris'' and in part by a CNRS PICS grant. G.K.L. would like to thank the  LPTHE in Paris and the ITP in Bern
  	for their kind hospitality while Y.C. would like to thank ITP, where part of the work was completed. Y.C. also thanks Taro Mori for help and useful discussions.

  \end{document}